\documentstyle[aps,prbbib,epsf,twocolumn,floats]{revtex}

        \newcommand{\beq}{\begin{eqnarray}}
        \newcommand{\eeq}{\end{eqnarray}}

        \newcommand{\bef}{\begin{figure}}
        \newcommand{\eef}{\end{figure}}
        
\begin{document}
\flushbottom

\wideabs{
\title{Unambiguous determination of the $g$-factor for holes in 
bismuth at high $B/T$}
\author{S. G. Bompadre$^a$, C. Biagini$^{a,b}$, D. Maslov$^a$
\& A. F. Hebard$^a$}
\address{$^a$ Department of Physics, University of
Florida, Gainesville, FL 32611-8440, USA\\
$^b$ Istituto Nazionale di Fisica della Materia,
L.go E. Fermi 2, 50125, Firenze, Italy}
\maketitle
\begin{abstract}
Magnetotransport has been investigated in high purity
bismuth crystals in static magnetic
fields as high as 20 T and temperatures as low as 25 mK.
This high $B/T$ ratio permits
observation of pronounced Shubnikov-de Haas oscillations over a
wide field range and
up to fields where most of the carriers are in the lowest
Landau level. For transport currents in the bisectrix or binary
directions and field in the perpendicular trigonal direction,
we have observed doublet splittings centered on each Shubnikov-de Haas
oscillation. These splittings exhibit a quadratic dependence on field and
disappear before the last oscillation. Our observations 
allow us to conclude unambiguously that already when the Landau
level index is as high as 2 the carriers are fully polarized and that
the $g$-factor for holes with the field in the trigonal
direction is 35.3(4).
\end{abstract}
}
\noindent
In this paper we report on magnetotransport measurements of
crystalline bismuth at high magnetic fields and low temperatures.
These measurements are motivated in part by the unusual electronic
properties of the semimetal bismuth that reflect its unique location
in an intermediate position between good metals and semiconductors.
The rhombohedral unit cell of bismuth can be obtained by slight
deformation of a double cubic cell and contains two atoms with five
electrons each. These ten electrons fill the first five Brillouin
zones and then spill into the sixth zone leaving behind hole pockets
in the fifth zone. The Fermi surface is well described by three
ellipsoids containing electrons lying almost in the bisectrix-binary plane
and one ellipsoid containing holes and oriented along the trigonal axis.
The small cross section of this highly anisotropic Fermi surface leads to a
small carrier density that is the same for both
electrons and holes ($n_{\rm hole}=n_{\rm electron}\approx
3\times10^{17}~{\rm cm}^{-3}$). Accordingly, bismuth is a perfectly
compensated semimetal. This unusual electronic structure together
with the ready availability of high quality crystals
with long mean free paths has generated a large field
of research \cite{Shoenberg1984,Edelmann1977}
on magnetic field induced oscillations in the magnetization
(de Haas-van Alphen effect) and conductivity (Shubnikov-de Haas effect).

For the highest magnetic fields, the carriers can be placed into
the lowest Landau level and the quantum limit realized. The direction of
the field is important. For example, if the field
is along the bisectrix direction (aligned closely
with the electron ellipsoids) then the quantum limit
can be realized at a few Tesla for the light electrons while the
heavier holes remain in the quasiclassical regime in fields up to 50 T.
Guided by these considerations and cognizant of the very limited
experimental evidence suggesting the existence of high magnetic-field-induced
correlated states,
\cite{Brandt1971,Yosida1976,Hiruma1979,Iye1982,Timp1983,Iye1984}
we have initiated magnetotransport studies of bismuth at high
$B/T$ with the aim of finding magnetic field induced
instabilities that might drive bismuth into a charge-density-wave (CDW)
insulating state. Although our initial results do not show
any evidence of such instabilities, they have clarified the meaning of doublet
structures that coincide with the Shubnikov-de Haas oscillations
and which have a heretofore-unobserved spacing that is quadratic in
field. The observations and analysis reported
here reveal an unambiguous determination of the $g$-factor
or equivalently the ratio of the Zeeman
energy to the cyclotron energy for holes confined to Landau orbits
by field directed along the trigonal axis. 

The bismuth crystals used in this study were obtained from a
variety of sources and had purities of 99.9995\%
or better. The crystals were cleaved at liquid nitrogen
temperatures to expose the trigonal plane
and then aligned along the bisectrix or binary directions according
to the striation marks created
during the low-temperature cleave. These alignments were confirmed
using Laue diffraction. Typical crystals were cut with a string saw to
5 mm length and had cross sectional areas on the order
of $4~{\rm mm}^2$. After cutting, the crystals were
etched in a nitric/acetic acid solution
and then annealed to temperatures as high as 250 C for
periods as long as a few days. The small area
contacts made to the samples using gold wires and low
melting temperature solder (Woods metal)
had contact resistances on the order of $0.1~\Omega$.
The current direction was chosen to be along
the binary (bisectrix) direction and the samples
were mounted on a rotating stage so that the applied
magnetic field could be rotated in the plane determined
by the trigonal and binary (bisectrix) directions.
All measurements were taken using low frequency ($<20~{\rm Hz}$) ac techniques.

Shown in Fig. (\ref{fig1}) is the dependence of
the longitudinal resistance on the reciprocal
magnetic field for a binary sample with the magnetic
field aligned perpendicular to the trigonal
plane. The temperature is at 25 mK and the magnetic
field was swept up to 20 T. This high $B/T$
allows more than 30 Shubnikov-de Haas oscillations
to be clearly discerned. For clarity, double
logarithmic axes were used to accommodate the greater
than two-decade range in field and greater
than three-decade range
in resistance. The choice of logarithmic axes however obscures the presence
of almost uniform spacing $\Delta(1/B)$ of the Shubnikov-de Haas oscillations.

Assigning consecutive integers $n$ to the resistance minima of
Fig. (\ref{fig1}) and then
plotting the values of $1/B$ at the respective minima versus $n$ addresses this
aspect of the data analysis. The plot of Fig. (\ref{fig2}) shows the results
of such a plot for a set of integers that are
consistent with a smooth extrapolation of the data to $n=0$. The set of
integers assigned in Fig. (\ref{fig2}) and labeled for $n=1$
in Fig. (\ref{fig1}) uniquely identifies the order of
each Shubnikov-de Haas oscillation. As will be shown below,
when spin splitting is taken into account, the order of oscillation
$n$ need not be the same as the the integer $\nu$ identifying the 
order of the Landau level.

At high fields the local slope in Fig. (\ref{fig2}) gives a
period for the Shubnikov-de Haas oscillations
$\Delta(1/B)=0.146~{\rm T}^{-1}$, in good agreement with experimental
values obtained by others for holes \cite{Smith1964,Brown1970} with the field
applied in the trigonal direction. The positive curvature reflects the fact
that the valence and conduction bands are affected by the field,
giving rise to a change in the carrier density and
thus in the Fermi energy. For the field along the trigonal axis both
the carrier density and the Fermi energy increase with the field.\cite{Smith1964} 
This field-induced increase in the Fermi energy and the
concomitant decrease in the $1/B$ spacings of the Shubnikov-de Haas
oscillations qualitatively explains the positive curvature. 
We note that this effect is actually quite small when compared to
what is expected when the field is applied in the basal
plane, where the Fermi energy has a stronger dependence on $B$.

 \bef
        \epsfxsize=10cm
        \centerline{\epsffile{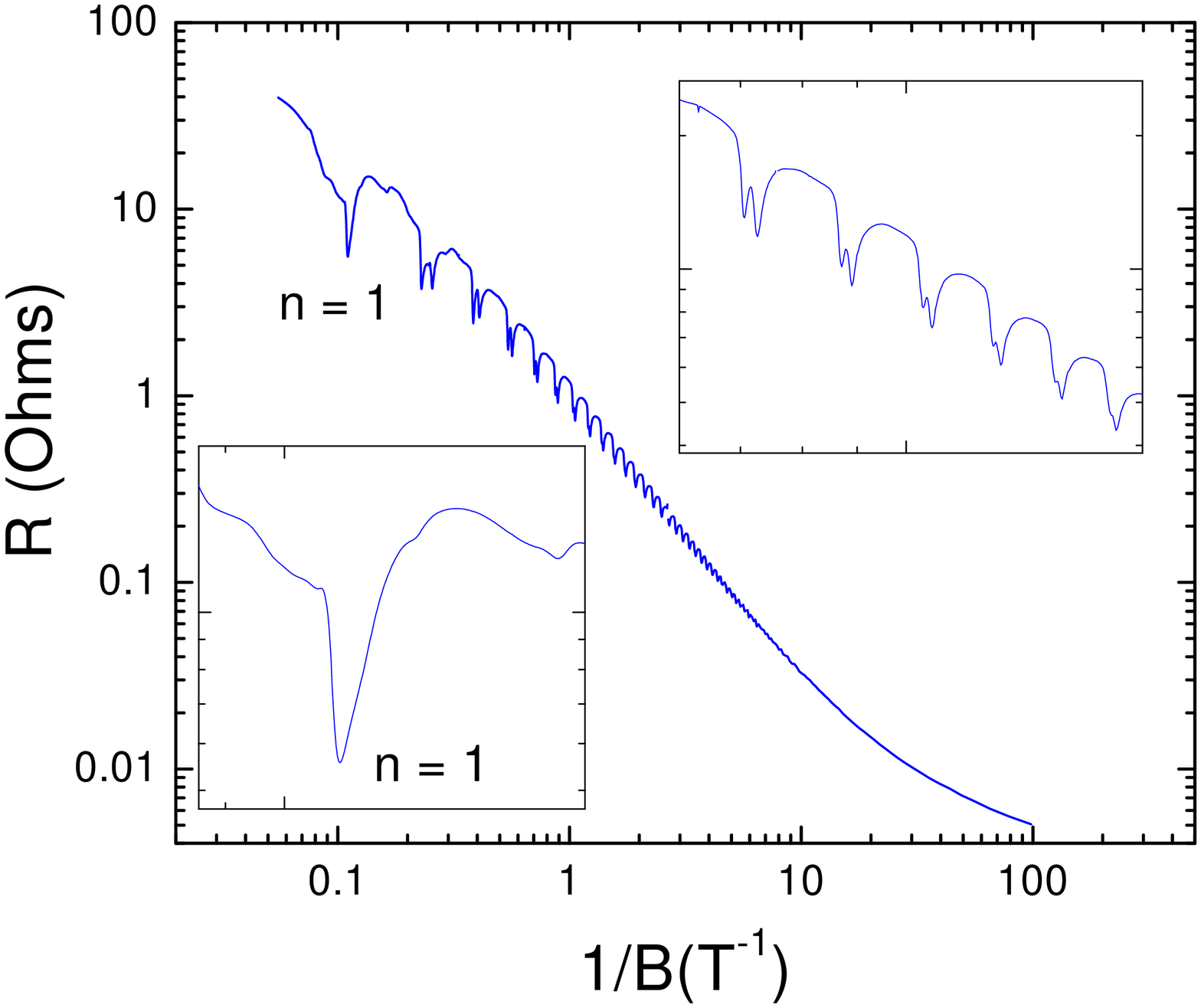}}
 \caption{Resistance vs reciprocal field on logarithmic axes. The
upper right inset is an expanded view of the $n=5$ through $n=10$
oscillations and the lower left inset is an expanded view of the
$n=1$ oscillation showing the sharpness of the resistance change
as the Fermi
energy passes through the fully polarized $\nu=1$ Landau level.}
 \label{fig1}
 \eef

 \bef
        \epsfxsize=10cm
        \centerline{\epsffile{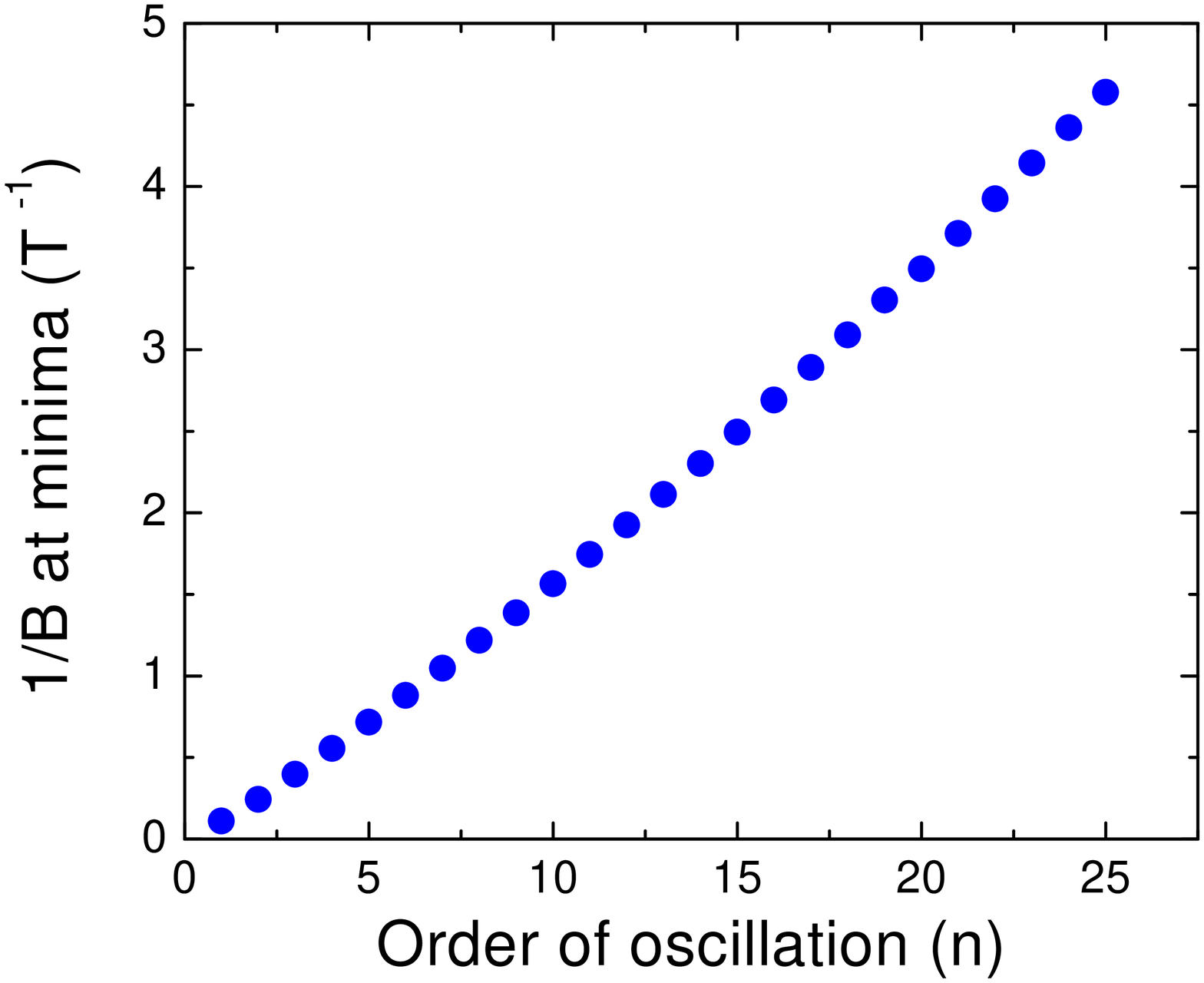}}
 \caption{Plot of the reciprocal fields as a function of integer $n$
for the first twenty-five
resistance minima
of Fig. 1.
The integers $n$ identify the order of the Shubnikov oscillation with $n=1$
labeled in Fig. 1}
 \label{fig2}
 \eef

A salient feature of the data presented in Fig. (\ref{fig1}) is the
appearance of doublets coinciding
with the valleys of the Shubnikov-de Haas oscillations. These doublets
persist to low fields as shown in the inset and as many as 10
can be clearly identified. In addition the spacing in the field
between the subpeaks of a doublet increases with the field. This
observation is made more quantitative in the plot of Fig. (\ref{fig3})
where the doublet spacings are plotted against the square of the field.

 \bef
        \epsfxsize=10cm
        \centerline{\epsffile{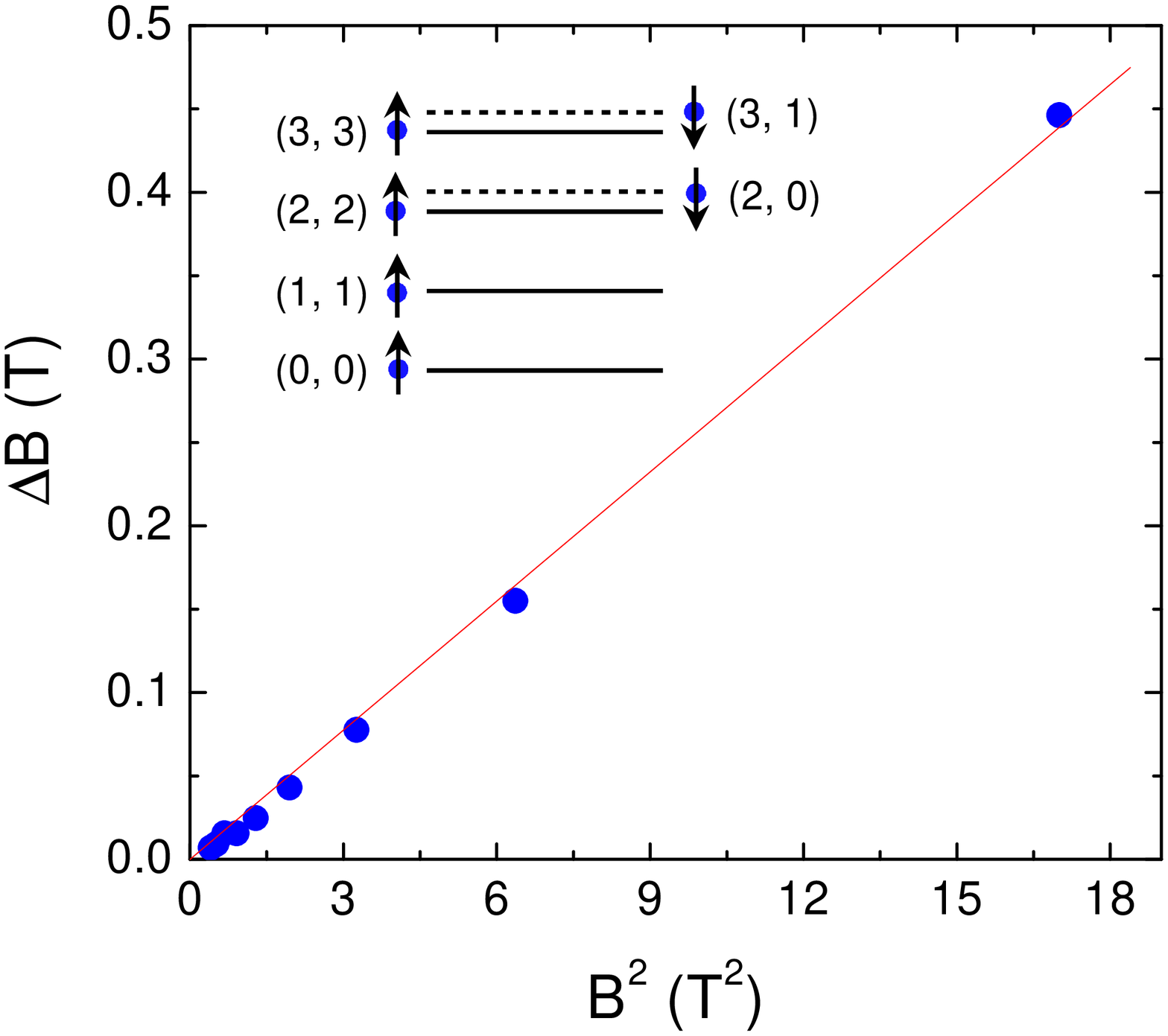}}
 \caption{Plot of the doublet spacings against the square of the field.
The inset is a schematic of the energy levels showing the spin
direction and the identifications $(n,\nu)$ of the order of oscillation
$n$ and Landau level $\nu$.}
 \label{fig3}
 \eef
 
For the oscillatory contribution to the holes' conductivity
for the field along the trigonal axis, one has \cite{LP79}
 \begin{eqnarray}
  \label{C9}
&&\tilde\sigma_{11}=\tilde\sigma_{22}\propto
        \frac{1}{B^{3/2}}\nonumber\\
&&\times\sum_{p=1}^\infty\!\frac{R_D(p)R_T(p)}{\sqrt p}
        \left\{\cos\left[2\pi p\left(
        \frac{l_B^2\epsilon_F}{\hbar^2\alpha_1}
        -\frac12-\frac g{2m_0\alpha_1}
        \right)-\frac\pi4 
        \right] \right. \nonumber \\
&&+ \left. \cos\left[2\pi p\left(
        \frac{l_B^2\epsilon_F}{\hbar^2\alpha_1}
        -\frac12+\frac g{2m_0\alpha_1}
        \right)-\frac\pi4\right]\right\}
\end{eqnarray}
where
 \beq
  \label{C6}
        R_T(p)=\frac{2\pi^2 p\frac{T l_B^2}
        {\hbar^2 \alpha_1}}{\sinh\left[
        2\pi^2 p\frac{T l_B^2}
        {\hbar^2 \alpha_1}\right]}
  \eeq
is the temperature smearing factor, and $R_D(p)$
is the disorder smearing factor.
The parameters $\alpha_i$ are the
components of the inverse mass tensor with respect to the
crystallographic axis (1 corresponds to
the binary axis, 2 to the bisectrix one and
3 to the trigonal axes: for the holes, $\alpha_1=\alpha_2$),
$m_0$ is the bare electron mass, $\epsilon_F\simeq0.012~{\rm eV}$
the hole Fermi energy at zero field, $g$ the hole gyromagnetic ratio and
$l_B=\sqrt{\hbar/eB}$ the magnetic length.
The disorder smearing factor
has the form
$$
R_D(T)=\exp\left[-\frac{2\pi^2 p T_Dl_B^2}{\hbar^2\alpha_1}\right]
$$
where $T_D=\hbar/2\pi\tau$ is 
the Dingle ``temperature'' and $\tau$ is the hole mean free time.
Using Eqn. (1) we analyzed the smearing of the Shubnikov-de Haas oscillations
at the relatively high temperature of 5K where the doublets are washed out
and the first harmonic ($p=1$) dominates.
At this temperature all of the broadening is found to be thermal, implying
that $T_D$ is well below 1K.
This estimate is consistent with our experimentally determined estimate
of $\tau$ ($\tau \simeq 2 \times 10^{-10}~{\rm s}$) and corresponding 
lower bound
on $T_D$ ($T_D \geq 0.006 {\rm K}$) calculated using the Drude formula 
and the measured resistivity at low temperature and zero field.

Eqn. (\ref{C9}) gives us the observed period with the magnetic
field.\cite{Shoenberg1984}
The signature of the crossing of a
Landau level by the
Fermi energy in a perfect crystal at zero temperature
is a {\em singularity} in the oscillations and a sharp peak in a real
crystal at high $B/T$ ratio.
In the experiment, we measured the resistivity but not the conductivity.
However, because bismuth is a perfectly compensated semimetal,
the Hall components of the conductivity tensor
are small compared to the diagonal ones. Therefore, the relation
between the conductivity and resistivity
 in a strong magnetic field is
very simple, namely
$\rho_{11}\approx1/\sigma_{11}$. 
Hence the oscillatory
corrections are related by
${\tilde\rho}_{11}/\rho_{11}=-{\tilde\sigma}_{11}/\sigma_{11}$. This means that the maxima
in ${\tilde\sigma}_{11}$, corresponding to the Fermi energy crossing
the center of the Landau level, appear as {\it minima} in
${\tilde\rho}_{11}$, as is observed experimentally (Fig. 1).

In  Eqn. (\ref{C9}) 
we emphasize the fact that the oscillatory behavior is given
by the superposition of two different oscillations, one for holes with
spin parallel to the field (``spin up'' holes) and the other for holes
with spin antiparallel to the field (``spin down'' holes).
As has already been pointed out, Eqn. (\ref{C9}) shows sharp features when the
argument of one of the cosines is not $p$-dependent;
in fact, in such a case
$\nu\propto\sum_{p=1}^\infty1/\sqrt p\to\infty$. This happens when
the factor multiplying $2\pi p$ is an integer number.
This singularity corresponds to the 1D singularity in the density of
states for the energy approaching the bottom of the Landau band.
Using this observation,
one readily finds
the spacing between resistance minima
for ``spin down'' and ``spin up'' holes taking the difference between
the position of two singularities
in Eqn. (\ref{C9})
with opposite
spin: this is given simply by
 \begin{equation}
\label{frac.g}
{\epsilon_F}\Delta\left(\frac1{\hbar\omega_c}\right)=
\frac{\epsilon_Z}{\hbar\omega_c}+\ell
 \end{equation}
where ${\epsilon_Z}$  is the Zeeman energy and $\ell$ is an arbitrary
integer number. The spacing in the inverse magnetic field is then given by
\begin{equation}
\Delta\left(\frac1B\right)=\frac{e\hbar}{\epsilon_F}\left(
\frac{g_0}{m_0}+\ell\alpha_1\right)
\end{equation}
$g_0$ is the part of the gyromagnetic ratio
which contributes only to the {\em fractional} part of the ratio between
the Zeeman energy and the cyclotron energy. The value of $g_0$
can then be obtained from the fit to the quadratic dependence in
Fig. (\ref{fig3}): $g_0 = 2.68(4)$.
The error here is determined by the scatter of the data.

At this point, we need to obtain the exact value of $\ell$, i.e.
the integer part of $\epsilon_Z/\hbar\omega_c$, which will be related
to a second contribution to the gyromagnetic ratio $g_1$:
usually, this is done from theoretical
considerations on the band structure
but in our case it is straightforward to
obtain this value from the observation that {\em the last
minimum} in Fig. (\ref{fig1}) is not split.
It is helpful to interpret
Eqn. (\ref{C9}) as the sum of two
oscillations with the same period, only
{\em slightly displaced in phase} due to the contribution
of the spin, in such a way that the first
peak of the doublet (in an increasing field) corresponds to spin down
holes and the second to spin up holes {\em with different Landau indices}.
The last split minimum at $n=2$ corresponds to the crossing
of the Fermi surface of the levels $(\nu=0,-)$ and $(\nu=2,+)$.
The spin
up holes are still oscillating: this is the meaning of the unsplit
minimum that signs the crossing through the Fermi surface
of the level $(\nu=1,+)$: therefore we can
conclude that the integer part of the ratio
of the Zeeman energy to the cyclotron
energy is equal to 2, which gives us $g_1=2m_0\alpha_1 = 32.6(3)$.
The precision in this evaluation is primarily due to uncertainties in
the effective mass determination.\cite{Shoenberg1984}
Summing up the two contributions, we obtain
\begin{equation}
g=g_1+g_0 = 35.3(4)
\end{equation}
which coincides with a ratio between the Zeeman energy to the cyclotron energy
equal to $\epsilon_Z/\hbar\omega_c = 2.16(2)$.

In agreement with previous investigations \cite{Brown1970}, no sign
of electronic contribution to the Shubnikov-de Haas
effect has been identified. The
electrons are expected to have a period three times
shorter than the holes, the difference being
due to the non parabolicity of their energy spectrum. At
the same time, the temperature and the Dingle
temperature smearing factor should have a bigger
numerical factor, in such a way that the electronic
oscillation should be depressed by a factor 3. But the
only signature of the presence of the electrons
is related to some modulation of the hole oscillation
that can be quantitatively interpreted
{\em assuming} a Dingle temperature for electrons
almost 10 times bigger than for the holes.
On the other hand, no theoretical arguments can be put
forward to explain the difference in the
Dingle temperatures viz. in the carrier mean free time.

In conclusion, we have shown in this work that the study of
magneto-oscillatory phenomena at high $B/T$ can reveal detailed
information about electronic structure that is not evident at higher
temperatures or lower fields. By high $B/T$ we mean fields high enough and
temperatures low enough such that $\hbar\omega_c\gg k_BT$.
In addition the system must be clean
enough, as are our bismuth crystals, to assure that the additional condition
$\hbar\omega_c\gg k_BT_D$ or,
equivalently $\omega_c\gg1/2\pi\tau$, holds. 
When both of these conditions are satisfied, higher
harmonics in Eqn. (\ref{C9}) become manifest
and quantitative information on the $g$-factor and spin polarization
can be obtained. With increasing field the
spin polarized Landau levels that cross the Fermi energy give rise to
increasingly sharp structures that are limited only by the disorder
broadened width of each Landau level. At $n=1$ near a field of $9~{\rm T}$
the sharpness is most pronounced (lower left insert of Fig. 1) with
a factor of two change in resistance for a less than 3\%
change in field.
We have shown that at high $B/T$ the appearance of
these sharp features is a direct consequence of the high harmonic
content of the
Shubnikov-de Haas oscillations which together with the
occurrence of well-defined spin-split doublets reveals
important details in electronic structure.

The authors would like to thank John Graybeal for his
interest and assistance in the initial
phase of this research and to F. Freibert for educating
us on the cutting and alignment of
bismuth crystals. The authors are also indebted to
R. Goodrich, Y. Liu, and C. Uher for the loan of high quality
bismuth crystals, without which this research could not have been
accomplished.
C.B. acknowledges support from the grant COFIN98-MURST and
 would also like to thank V. Kotov and G. Martin for many valuable
discussions.
D.L.M. acknowledges the support from the Research Corporation
Innovation Award RI0082.
This work was supported by the NSF funded In-House Research Program
(IHRP) of the NHMFL in Tallahassee and the authors are particularly
appreciative of the help provided at that facility by D. Hall,
T. Murphy and
E. Palm of the Users Support Group.

\end{document}